A. V. Goryunov 

# ИДУЩАЯ ВОЛНА КАК МОДЕЛЬ ЧАСТИЦЫ

А.В. Горюнов

С классических релятивистских позиций введено понятие *идущей* волны. Рассмотрены одномерная и трёхмерная идущие волны, их волновые и дисперсионные уравнения. Показано, что волновые (длины волн де Бройля и Комптона) и корпускулярные (4-вектор энергии-импульса и *масса покоя*) характеристики микрочастицы могут быть выражены через параметры идущей волны. Тем самым, предложен новый взгляд на ряд уже устоявшихся понятий физики, связанных с концепцией корпускулярно-волнового дуализма.

Концепция корпускулярно-волнового дуализма является одним из базовых понятий современной квантовой теории. Само возникновение этой концепции было обусловлено не только наличием волновых и корпускулярных свойств у микрообъектов, но и отсутствием единого способа их описания в рамках уже имевшихся классических понятий. Так, идея чисто волнового описания микрочастиц возникла одновременно с появлением работ де Бройля и Шрёдингера по «волновой» механике. Последующий отказ от такого подхода так же имел своё обоснование. С одной стороны, это «нематериальность» волн де Бройля: их фазовая скорость превосходит скорость света, и физический смысл удавалось придать только групповой скорости – именно она совпадает со скоростью движения частицы. С другой стороны, волновой пакет, образованный группой волн де Бройля, распространяющихся в одном направлении, не является стабильным (он расплывается) и идея связывания локализации частицы с максимумом амплитуды волнового пакета теряет смысл. Вероятностное истолкование волновой функции наряду с успехами и внутренней самосогласованностью квантовой механики в целом, постепенно перевели классическую физику в ранг устаревшей теории, как бы в принципе непригодной для адекватного описания микрообъектов.

В данной работе с классических релятивистских позиций введено понятие *идущей* волны. Анализ различных аспектов этого понятия проведён на нескольких простейших теоретических моделях. Показано, что волновые и корпускулярные характеристики микрочастицы могут быть выражены через параметры идущей волны. Тем самым, предложен новый взгляд на ряд уже устоявшихся понятий физики, связанных с концепцией корпускулярно-волнового дуализма.

### 1. Времениподобный вектор как сумма изотропных векторов

В двухмерном $(t, x)$ пространстве-времени (рис.1) имеется два (пространственно противоположных) изотропных направления. При сложении коллинеарных изотропных векторов получим изотропный вектор того же направления, например $\mathbf{b} + \alpha\mathbf{b} = (1 + \alpha)\mathbf{b}$, где $\alpha$ – произвольное действительное число. Если же складывать

А.В. Горюнов

пространственно противоположные изотропные векторы $\mathbf{b} = (t_b, x_b)$ и $\mathbf{a} = (t_a, -x_a)$, то в результате получим уже не изотропный, а времениподобный вектор $\mathbf{c} = \mathbf{b} + \mathbf{a} = (t_b + t_a,\ x_b - x_a) = (t_c,\ x_c)$. Верно и обратное утверждение: любой (двухмерный) времениподобный вектор может быть однозначно разложен на два изотропных слагаемых. Квадрат интервала вектора $\mathbf{c}$ выражается через координаты векторов $\mathbf{a}$ и $\mathbf{b}$ следующим образом:

$$s^2 = c^2 t_c{}^2 - x_c{}^2 = c^2(t_b + t_a)^2 - (x_b - x_a)^2 =$$
$$= (c^2 t_a{}^2 - x_a{}^2) + (c^2 t_b{}^2 - x_b{}^2) + 2(c^2 t_a t_b + x_a x_b)$$

В правой части этого выражения первые два слагаемых равны нулю как квадраты интервалов изотропных векторов $\mathbf{a}$ и $\mathbf{b}$. Поэтому

(1.1)
$$s^2 = c^2 t_c{}^2 - x_c{}^2 = 2(c^2 t_a t_b + x_a x_b)$$

В (1.1) правая часть есть удвоенное псевдоевклидово скалярное произведение векторов $\mathbf{a}$ и $\mathbf{b}$, которое при преобразованиях Лоренца сохраняется. Отметим, что отличие вида этого скалярного произведения от стандартного ($c^2 t_1 t_2 - x_1 x_2$) в данном случае связано просто с явным учётом знаков координат в обозначениях:

$$c^2 t_a t_b - (-x_a) x_b = c^2 t_a t_b + x_a x_b$$

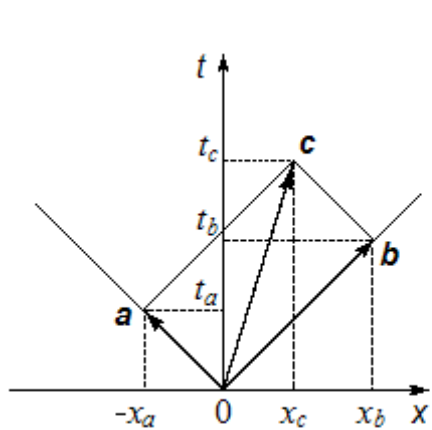 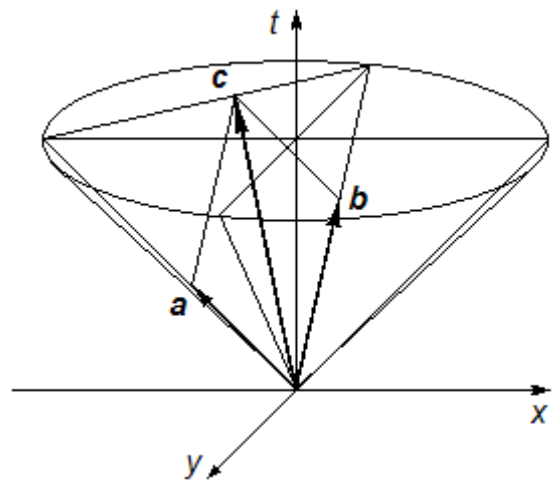

Рис. 1             Рис. 2

В 3-мерном $(t, x, y)$ или в 4-мерном $(t, x, y, z)$ пространстве-времени, для получения времениподобного вектора, складываемые изотропные векторы уже не обязаны быть пространственно противоположными – достаточно, чтобы они не были коллинеарными (рис. 2). Таким образом, и в 4-мерном пространстве-времени, сложение любых двух неколлинеарных изотропных 4-векторов (будущего) даёт времениподобный 4-вектор с ненулевой псевдоевклидовой нормой, вычисляемой аналогично (1.1):

$$s^2 = c^2 t_c{}^2 - x_c{}^2 = 2(c^2 t_a t_b - x_a x_b - y_a y_b - z_a z_b) \qquad (1.2)$$

В частности, это будет относиться и к 4-вектору энергии-импульса, $(E/c,\ p_x, p_y, p_z)$, норма которого, $m_0 c$, пропорциональна *массе покоя*. Тем самым, наличие массы покоя у частицы может рассматриваться как результат сложения некоторых безмассовых изотропных процессов.

## 2. Суперпозиция встречных волн как модель частицы

Рассмотрим стоячую волну, образованную в системе отсчёта $O'(t', x')$ суперпозицией двух встречных бегущих волн, одинаковых по частоте и амплитуде:

(2.1)
$$\cos(\omega_0 t' - k_0 x') + \cos(\omega_0 t' + k_0 x') = 2\cos(k_0 x')\cos(\omega_0 t')$$

Скорость бегущих волн равна скорости света в вакууме $c$, а $\omega_0 = c k_0$. В правой части этого выражения сомножитель $2|\cos(k_0 x')|$ обычно интерпретируют как (зависящую от $x'$) амплитуду колебаний, в свою очередь зависящих от времени по закону $\cos(\omega_0 t')$.

В системе отсчёта $O(t, x)$, движущейся относительно исходной системы со скоростью $v$ навстречу первой волне, частоты обеих волн изменятся согласно эффекту Доплера:

(2.2)
$$\omega_1 = \omega_0 \sqrt{\frac{1 + v/c}{1 - v/c}} = \omega_0 \frac{1 + v/c}{\sqrt{1 - (v/c)^2}}, \qquad \omega_2 = \omega_0 \sqrt{\frac{1 - v/c}{1 + v/c}} = \omega_0 \frac{1 - v/c}{\sqrt{1 - (v/c)^2}},$$

а координаты $(t, x)$ и $(t', x')$ связаны преобразованием Лоренца. Запишем суперпозицию (2.1) в системе $O$ в следующем виде:

(2.3)
$$\cos(\omega_1 t - k_1 x) + \cos(\omega_2 t + k_2 x) = 2\cos(\omega t - Kx)\cos(\Omega t - kx),$$

где введены обозначения:

(2.4)
$$\frac{\omega_1 + \omega_2}{2} = \Omega, \qquad \frac{\omega_1 - \omega_2}{2} = \omega, \qquad \frac{k_1 + k_2}{2} = K, \qquad \frac{k_1 - k_2}{2} = k.$$

Характерные особенности (например, узлы) волны теперь уже не будут стоять на месте, но будут перемещаться в системе $O$ со скоростью, зависящей от её движения по отношению к исходной системе $O'$. Будем называть волну (2.3) – *идущей*, отмечая тем самым её отличие от *стоячей* волны (2.1) и образующих её встречных *бегущих* волн.

В системе отсчёта $O$ обе бегущие волны по-прежнему распространяются со скоростью света $c$ и для них выполняются соотношения $\omega_1 = c k_1$ и $\omega_2 = c k_2$, тогда как для величин (2.4) получим:

$$\frac{\Omega}{k} = \frac{\omega_1 + \omega_2}{k_1 - k_2} = V > c, \qquad \Omega = Vk = cK, \qquad \frac{2\pi}{k} = \Lambda, \qquad (2.5)$$

$$\frac{\omega}{K} = \frac{\omega_1 - \omega_2}{k_1 + k_2} = v < c, \qquad \omega = vK = ck, \qquad \frac{2\pi}{K} = \lambda. \qquad (2.6)$$

Формула (2.3) выражает суперпозицию встречных волн в виде произведения двух волновых сомножителей – высокочастотного ($\Omega$) длинноволнового ($\Lambda$), распространяющегося со сверхсветовой скоростью $V$, и низкочастотного ($\omega$) коротковолнового ($\lambda$), распространяющегося со скоростью $v < c$. Если сомножитель $2|\cos(\omega t - Kx)|$ по-прежнему интерпретировать как амплитуду колебаний, то теперь эта амплитуда не только распределена в пространстве определённым (зависящим от $x$) образом, но это распределение смещается вдоль оси $x$ с постоянной скоростью $v$. В свою очередь, колебания с частотой $\Omega$ зависят не только от времени $t$, но и от координаты точки $x$ по закону $\cos(\Omega t - kx)$, что выглядит как распространение волны со скоростью $V$.

Согласно (2.5) и (2.6), скорости и частоты волн-сомножителей соотносятся следующим образом:

$$vV = \frac{\omega \Omega}{Kk} = c^2, \qquad V = \frac{c^2}{v}, \qquad v = \frac{c^2}{V}, \qquad \frac{v}{c} = \frac{c}{V} = \frac{\omega}{\Omega}. \qquad (2.7)$$

Подчеркнём, что скорость низкочастотной волны-сомножителя и скорость системы отсчёта не случайно обозначены одной и той же буквой $v$. Так, в (2.7) $v$ входит как скорость волны. Если же частоты $\Omega$ и $\omega$ выразить на основе (2.2) через скорость системы отсчёта, то получим

$$\Omega = \frac{\omega_1 + \omega_2}{2} = \omega_0 \frac{1 + v/c + 1 - v/c}{2\sqrt{1 - (v/c)^2}} = \omega_0 \frac{1}{\sqrt{1 - (v/c)^2}}, \qquad (2.8)$$

$$\omega = \frac{\omega_1 - \omega_2}{2} = \omega_0 \frac{1 + v/c - 1 + v/c}{2\sqrt{1 - (v/c)^2}} = \omega_0 \frac{v/c}{\sqrt{1 - (v/c)^2}}, \qquad (2.9)$$

откуда следует, что и в этом случае $\omega/\Omega = v/c$.

Будучи образованной двумя бегущими волнами, имеющими в любой системе отсчёта световую скорость и, следовательно, не имеющих массы покоя, сама идущая волна ведёт себя как объект с ненулевой массой покоя. Так,

$$\Omega + \omega = \omega_1 = \omega_0 \sqrt{\frac{1 + v/c}{1 - v/c}}, \qquad \Omega - \omega = \omega_2 = \omega_0 \sqrt{\frac{1 - v/c}{1 + v/c}}, \qquad (2.10)$$

а произведение этих выражений равно

$$(2.11)$$

А.В. Горюнов

$$\Omega^2 - \omega^2 = \omega_1\omega_2 = {\omega_0}^2,$$

что естественно сопоставляется с релятивистским уравнением

$$E^2 - (pc)^2 = (m_0 c^2)^2$$

и базовыми квантовыми соотношениями («~» обозначает пропорциональность)

(2.12)

$$\hbar\omega_0 \sim m_0 c^2,$$

$$\hbar\Omega = \frac{\hbar\omega_0}{\sqrt{1-(v/c)^2}} \sim \frac{m_0 c^2}{\sqrt{1-(v/c)^2}} = mc^2 = E,$$

$$\hbar\omega = \frac{\hbar\omega_0\, v/c}{\sqrt{1-(v/c)^2}} \sim \frac{m_0 vc}{\sqrt{1-(v/c)^2}} = mvc = pc,$$

где $m = m_0/\sqrt{1-(v/c)^2}$, а $\omega_0$ – та самая частота внутреннего колебательного процесса, присущего частице в смысле де Бройля.

При $\omega_1 \to \omega_0$ и $\omega_2 \to \omega_0$ ($\omega_1 > \omega_0 > \omega_2$) из (2.5) и (2.6) следует, что

$$\begin{aligned}
\Omega &\to \omega_0, & k &\to 0, & \Lambda &\to \infty, & V &\to \infty, \\
\omega &\to 0, & K &\to k_0, & \lambda &\to \lambda_0, & v &\to 0,
\end{aligned}$$

и соотношение (2.3) опять примет вид (2.1). Поэтому при совпадении частот стоячая волна, все точки которой колеблются синфазно, формально может рассматриваться как волна, фаза которой мгновенно ($V = \infty$) распространяется на бесконечно большое расстояние ($\Lambda = \infty$); амплитуда же колебаний каждой точки – неизменна, т. е. скорость распространения изменения амплитуды $v = 0$.

Бесконечная скорость распространения фазы не противоречит теории относительности, так как при этом не происходит переноса энергии вдоль стоячей волны – встречный перенос энергии бегущими волнами взаимно скомпенсирован. Все бесконечности здесь возникают в связи с тем, что рассматриваемый объект – гармоническая волна – можно сказать, изначально бесконечен, как во времени, так и в пространстве.

При другой крайней ситуации, когда $v \to c$, из (2.2) получим $\omega_1 \to \infty$ и $\omega_2 \to 0$, а из (2.5) и (2.6)

$$\omega \to \Omega \to \infty, \quad k \to K \to \infty, \quad \Lambda \to \lambda \to 0 \quad \text{и} \quad V \to c.$$

Тем самым, имеющая массу покоя идущая волна приближается по своим свойствам к высокочастотной бегущей волне.

Если соотношения (2.12) рассматривать как точные равенства, то с учётом (2.5) и (2.6) из них следует

А.В. Горюнов

$$mv = \hbar k, \qquad mc = \hbar K \qquad (2.13)$$

и

$$\frac{2\pi\hbar}{mv} = \Lambda, \qquad \frac{2\pi\hbar}{mc} = \lambda \qquad (2.14)$$

то есть, $\Lambda$ соответствует длине волны де Бройля, а $\lambda$ – длина волны, которая в пределе стоячей волны ($v \to 0$, $\lambda \to \lambda_0$, $m \to m_0$) совпадает с комптоновской: $2\pi\hbar/m_0 c = \lambda_0$.

Отметим, что уменьшение длины волны де Бройля $\Lambda$ при увеличении скорости $v$ никак не связано с уточнением локализации частицы – «частицей» является вся идущая волна, бесконечная как в пространстве, так и во времени.

Согласно (2.12), масса покоя частицы $m_0 \sim \omega_0$, а по формуле (2.11) $\omega_0^2 = \omega_1\omega_2$. То есть, неизменность массы покоя частицы при изменении её скорости обеспечивается взаимообратным изменением частот $\omega_1$ и $\omega_2$ суммирующихся встречных бегущих волн.

Стоячая и идущая волны получаются при сложении встречных волн *одинаковой амплитуды*. Если амплитуды различны, то «излишек» ($a$) остаётся у бегущей волны и в образовании стоячей и идущей волн участия не принимает:

(2.15)
$$(A + a)\cos(\omega_0 t - k_0 x) + A\cos(\omega_0 t + k_0 x) = a\cos(\omega_0 t - k_0 x) + 2A\cos(k_0 x)\cos(\omega_0 t)$$
$$(A + a)\cos(\omega_1 t - k_1 x) + A\cos(\omega_2 t + k_2 x) =$$
$$= a\cos(\omega_1 t - k_1 x) + 2A\cos(\omega t - Kx)\cos(\Omega t - kx)$$

### 3. Модель неупругого взаимодействия

Рассмотрим две идущие волны, которые в соответствии с (2.1), (2.4) и (2.11) описываются следующими соотношениями:

$$\Omega_1 = (\omega_{11} + \omega_{12})/2, \qquad ck_1 = (\omega_{11} - \omega_{12})/2, \qquad \Omega_1^2 - (ck_1)^2 = \omega_{11}\omega_{12} = \omega_{10}^2,$$
$$\Omega_2 = (\omega_{21} + \omega_{22})/2, \qquad ck_2 = (\omega_{21} - \omega_{22})/2, \qquad \Omega_2^2 - (ck_2)^2 = \omega_{21}\omega_{22} = \omega_{20}^2.$$

Идущая волна 1 образована встречными бегущими волнами 11 и 12, а идущая волна 2 – встречными бегущими волнами 21 и 22. Будем считать, что бегущие волны 11 и 21 распространяются по направлению оси $Ox$, а волны 12 и 22 – против. Взаимодействие состоит в «переключении» встречных бегущих волн от одной идущей волны к другой. В результате, идущие волны 1 и 2 исчезают, и возникают идущие волны 3 и 4:

$$\Omega_3 = (\omega_{11} + \omega_{22})/2, \qquad ck_3 = (\omega_{11} - \omega_{22})/2, \qquad \Omega_3^2 - (ck_3)^2 = \omega_{11}\omega_{22} = \omega_{30}^2,$$
$$\Omega_4 = (\omega_{21} + \omega_{12})/2, \qquad ck_4 = (\omega_{21} - \omega_{12})/2, \qquad \Omega_4^2 - (ck_4)^2 = \omega_{21}\omega_{12} = \omega_{40}^2.$$

Механизм объединения четырёх бегущих волн в конкретные две пары, составляющие идущие волны, как и механизм их «переключения», остаётся за рамками данной модели. Сама же модель иллюстрирует принцип превращения одних взаимодейст-вующих частиц в другие при сохранении 4-вектора энергии-импульса для системы в целом. Так, с учётом соотношений (2.12), имеем

$$\Omega_1 + \Omega_2 = (\omega_{11} + \omega_{12} + \omega_{21} + \omega_{22})/2 = \Omega_3 + \Omega_4 \to E_1 + E_2 = E_3 + E_4$$
$$ck_1 + ck_2 = (\omega_{11} - \omega_{12} + \omega_{21} - \omega_{22})/2 = ck_3 + ck_4 \to p_1 + p_2 = p_3 + p_4$$

при этом частицы с массами покоя $m_{10} \sim \omega_{10}$ и $m_{20} \sim \omega_{20}$ превращаются в частицы с массами покоя $m_{30} \sim \omega_{30}$ и $m_{40} \sim \omega_{40}$. Например, при встречном столкновении быстрых частиц, для которых выполняются условия $\omega_{11} \gg \omega_{12}$ и $\omega_{21} \ll \omega_{22}$, возникает частица с массой $m_{30} \sim \omega_{30} = (\omega_{11}\omega_{22})^{1/2}$, превосходящей массы покоя исходных частиц, $m_{10}$ и $m_{20}$. Если (как данная модель и предполагает) массу покоя имеют все начальные и конечные участники реакции, то эти массы связаны соотношением:

$$m_{10}\, m_{20} = m_{30}\, m_{40} \qquad (3.1)$$

поскольку $(\omega_{10}\, \omega_{20})^2 = (\omega_{11}\, \omega_{12}\, \omega_{21}\, \omega_{22}) = (\omega_{30}\, \omega_{40})^2$.

### 4. Устойчивые суперпозиции идущих волн

Дисперсией волн называют зависимость фазовой скорости гармонической волны от её частоты. Закон дисперсии обычно задают в виде дисперсионного уравнения, связывающего частоту и волновой вектор гармонической волны: $\omega = \omega(k)$. Примером недиспергирующих волн являются электромагнитные волны в вакууме, для которых выполняется дисперсионное уравнение $\omega/k = c = \text{const}$. Однонаправленные недиспергирующие волны могут сформировать устойчивый волновой пакет, скорость движения которого равна фазовой скорости образующих его бегущих волн. При наличии же дисперсии, разночастотные составляющие волнового пакета, распространяясь с различной скоростью, постепенно «растащат» свои максимумы и волновой пакет расплывётся.

Рассматриваемые нами стоячие и идущие волны являются суммой бегущих волн, распространяющихся с постоянной – световой – скоростью, для которых дисперсия отсутствует. Представление суммы недиспергирующих волн в виде произведения волн, скорость которых зависит от частоты, даёт возможность взглянуть на ситуацию с новой точки зрения, но не меняет её сущности. Так, для стоячей волны (2.1) с другой базовой частотой $\omega_0'$ будут получены и другие частоты волн-сомножителей (2.3) $\Omega'$ и $\omega'$, но их отношение в одной и той же системе отсчёта будет одинаковым: $\omega/\Omega = \omega'/\Omega' = v/c$ и определяться оно будет именно скоростью движения этой системы. Несмотря на наличие дисперсии волн-сомножителей, частоты различных идущих волн связаны со своими волновыми векторами посредством одних и тех же скоростей ($0 < v < c$ и $c < V < \infty$):

$$\Omega = Vk, \quad \omega = vK \quad \text{и} \quad \Omega' = Vk', \; \omega' = vK'. \qquad (4.1)$$

Таким образом, суперпозиция, образованная наложением различных стоячих (в одной и той же системе отсчёта) волн, не будет расплываться и в любой другой системе отсчёта, движущейся с произвольной скоростью $v < c$. Отметим, что базовые частоты $\omega_0$ и $\omega_0'$ (или волновые векторы $k_0$ и $k_0'$) стоячих волн, образующих устойчивую суперпозицию, совсем не обязаны быть близкими по величине. Примером такой суперпозиции может служить сумма двух (или нескольких) стоячих волн с кратной длиной волны: их общие узлы останутся общими и в системе отсчёта идущей волны.

### 5. Волновые и дисперсионные уравнения

Известно, что дисперсионное уравнение $\omega^2 - c^2 k^2 = (m_0 c^2/\hbar)^2$ соответствует релятивистскому волновому уравнению Клейна – Гордона

$$\psi_{tt} = c^2 \psi_{xx} - (m_0 c^2/\hbar)^2 \psi \tag{5.1}$$

для гармонической волны де Бройля $\psi = A\exp[-i(\omega t \pm kx)]$. Эти дисперсионное и волновое уравнения описывают свободную частицу с массой покоя $m_0$. Обычно подчёркивается, что при $m_0 = 0$ эти уравнения переходят в классические уравнения для волн, распространяющихся со световой скоростью: $\omega = ck$ и $\psi_{tt} = c^2 \psi_{xx}$.

Запишем соотношение для идущей волны (2.3) в следующей краткой форме:

$$f = f_1 + f_2 = 2\varphi\Phi \tag{5.2}$$

Вторые частные производные $f$ по времени и по пространственной координате обозначим нижними индексами $tt$ и $xx$, а штрихами – производные функций по собственным фазам:

$$f_{tt} = \omega_1{}^2 f_1'' + \omega_2{}^2 f_2'' = 2(\omega^2 \varphi''\Phi + 2\omega\Omega\, \varphi'\Phi' + \Omega^2 \varphi\Phi'')$$
$$f_{xx} = k_1{}^2 f_1'' + k_2{}^2 f_2'' = 2(K^2 \varphi''\Phi + 2Kk\, \varphi'\Phi' + k^2 \varphi\Phi'')$$

В суммарной форме $(f_1 + f_2)$ функция $f$ удовлетворяет волновому уравнению

$$f_{tt} = c^2 f_{xx} \tag{5.3}$$

в соответствии с законом дисперсии отдельных слагаемых:

$$\omega_1 = ck_1, \qquad \omega_2 = ck_2. \tag{5.4}$$

Тому же волновому уравнению, естественно, удовлетворяет и $\varphi\Phi$-форма функции $f$:

$$\omega^2 \varphi''\Phi + 2\omega\Omega\, \varphi'\Phi' + \Omega^2 \varphi\Phi'' = c^2(K^2 \varphi''\Phi + 2Kk\, \varphi'\Phi' + k^2 \varphi\Phi'')$$

Сократим обе части уравнения на одинаковые выражения $2\omega\Omega\ \varphi'\Phi' = 2c^2 Kk\ \varphi'\Phi'$, а оставшиеся слагаемые сгруппируем:

$$(\Omega^2 - c^2 k^2)\varphi\Phi'' = (c^2 K^2 - \omega^2)\varphi''\Phi$$

поскольку $(c^2 K^2 - \omega^2) = (\Omega^2 - \omega^2) = (\Omega + \omega)(\Omega - \omega) = \omega_1 \omega_2 = \omega_0{}^2$, а $\varphi\Phi'' = \varphi''\Phi = -\varphi\Phi$, то

$$\Omega^2 - c^2 k^2 = \omega_0{}^2 \qquad (5.5)$$

Таким образом, дисперсионные уравнения (5.4) и (5.5) не противоречат друг другу и оба соответствуют обычному волновому уравнению (5.3) для идущей волны (2.3).

## 6. Сферическая волна

Стоячую сферическую волну с *отсутствующим источником* можно представить так: некоторый процесс распространяется по радиусам от периферии к центру, проходит через этот центр, и далее по противоположным радиусам (тех же диаметров) продолжает распространяться от центра к периферии. Тем самым, один и тот же, непрерывный вдоль диаметра процесс сначала выступает в роли сходящейся волны, а потом – после прохождения центра – в роли волны расходящейся. Таким образом, по каждому диаметру одновременно идут два встречных процесса. В результате, вдоль каждого диаметра формируется одномерная стоячая волна, а в пространстве – 3-мерная сферическая стоячая волна.

Эту же ситуацию можно интерпретировать и иначе: источник расходящейся волны и сток сходящейся – равны по величине, расположены в одной точке (центре сферы) и взаимно компенсируют друг друга. Сумма таких волн имеет следующий (формально одномерный по $R$) вид:

$$\frac{1}{R'}\cos(\omega_0 t' - k_0 R') + \frac{1}{R'}\cos(\omega_0 t' + k_0 R') = \frac{2}{R'}\cos(k_0 R')\cos(\omega_0 t'). \qquad (6.1)$$

На данном этапе нас интересуют не амплитудные, а фазовые закономерности. Поэтому далее будем использовать выражение (6.1), сократив его на общий множитель $(1/R')$. Если перейти в движущуюся относительно $O'$ систему отсчёта $O$, то стоячая в $O'$ волна станет в $O$ идущей. Направим ось $x$ системы отсчёта $O$ по движению идущей волны. Идущая волна утратит присущую стоячей волне сферическую симметрию, но сохранит аксиальную относительно оси $Ox$. Поэтому в качестве второй координаты удобно выбрать радиус $r$, перпендикулярный оси $Ox$ (рис. 3).

А.В. Горюнов

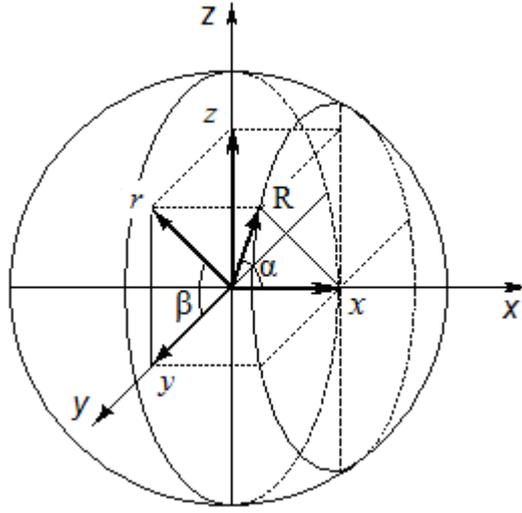

Рис. 3.

Тем самым, от одномерной формы записи (6.1) сферической стоячей волны мы перейдём к её двухмерной форме:

(6.2)
$$\cos(\omega_0 t' - k_{0x} x' - k_{0r} r') + \cos(\omega_0 t' + k_{0x} x' + k_{0r} r') = 2\cos(k_{0x} x' + k_{0r} r')\cos(\omega_0 t'),$$
$$\cos(\omega_0 t' - k_{0x} x' + k_{0r} r') + \cos(\omega_0 t' + k_{0x} x' - k_{0r} r') = 2\cos(k_{0x} x' - k_{0r} r')\cos(\omega_0 t'),$$

(6.3)
где $x' = R' \cos\alpha$, $r' = R' \sin\alpha$, $k_{0x} = k_0 \cos\alpha$, $k_{0r} = k_0 \sin\alpha$ и $0 \leq \alpha \leq \pi/2$. (6.4)

Формула (6.2) описывает стоячую волну в передней полусфере, когда проекции волнового вектора **k** на оси $x$ и $r$ одного знака, а формула (6.3) – в задней полусфере, когда знаки проекций вектора **k** на эти оси противоположны.

При переходе в систему $O$, для всех 4-векторов $t$- и $x$-составляющие преобразуются согласно формулам Лоренца; $y$- и $z$-, а значит и $r$-составляющие, не изменяются. Соответствующим образом преобразуется и 4-вектор $(\omega_0/c,\ \mathbf{k}_0)$:

$$\frac{\omega_1}{c} = \frac{\omega_0/c + (v/c) k_{0x}}{\sqrt{1-(v/c)^2}}, \qquad k_{1x} = \frac{k_{0x} + (v/c)(\omega_0/c)}{\sqrt{1-(v/c)^2}}, \qquad k_{1r} = k_{0r},$$

$$\frac{\omega_2}{c} = \frac{\omega_0/c - (v/c) k_{0x}}{\sqrt{1-(v/c)^2}}, \qquad k_{2x} = \frac{k_{0x} - (v/c)(\omega_0/c)}{\sqrt{1-(v/c)^2}}, \qquad k_{2r} = k_{0r},$$

Учитывая, что $\omega_0 = c k_0$ и формулы (6.4), получим:

(6.5)
$$\omega_1 = \omega_0 \frac{1 + (v/c)\cos\alpha}{\sqrt{1-(v/c)^2}} = c k_1, \qquad k_{1x} = k_0 \frac{\cos\alpha + (v/c)}{\sqrt{1-(v/c)^2}}, \qquad k_{1r} = k_0 \sin\alpha,$$

$$\omega_2 = \omega_0 \frac{1 - (v/c)\cos\alpha}{\sqrt{1-(v/c)^2}} = c k_2, \qquad k_{2x} = k_0 \frac{\cos\alpha - (v/c)}{\sqrt{1-(v/c)^2}}, \qquad k_{2r} = k_0 \sin\alpha,$$

(6.6)

что представляет собой одну из форм записи трёхмерного (формально – двухмерного) эффекта Доплера, удобную именно для рассматриваемой здесь ситуации и переходящую в одномерные формулы (1.2) при $\alpha = 0$. Используя (6.5) и (6.6), прямым вычислением можно убедиться, что

$$k_{1x}^2 + k_{1r}^2 = k_1^2 \quad \text{и} \quad k_{2x}^2 + k_{2r}^2 = k_2^2 \tag{6.7}$$

и тем самым, что 4-векторы $(\omega_1/c, \mathbf{k}_1)$ и $(\omega_2/c, \mathbf{k}_2)$ – изотропные.

В системе $O$ соотношения для трёхмерной идущей волны примут вид:

$$\cos(\omega_1 t - k_{1x}x - k_{1r}r) + \cos(\omega_2 t + k_{2x}x + k_{2r}r) = 2\cos(\omega t - K_x x - K_r r)\cos(\Omega t - k_x x) \tag{6.8}$$

$$\cos(\omega_1 t - k_{1x}x + k_{1r}r) + \cos(\omega_2 t + k_{2x}x - k_{2r}r) = 2\cos(\omega t - K_x x + K_r r)\cos(\Omega t - k_x x) \tag{6.9}$$

где

$$\Omega = \frac{\omega_1 + \omega_2}{2} = \omega_0 \frac{1}{\sqrt{1-(v/c)^2}}, \qquad K_x = \frac{k_{1x} + k_{2x}}{2} = k_0 \frac{\cos\alpha}{\sqrt{1-(v/c)^2}}$$

$$\omega = \frac{\omega_1 - \omega_2}{2} = \omega_0 \frac{(v/c)\cos\alpha}{\sqrt{1-(v/c)^2}}, \qquad k_x = \frac{k_{1x} - k_{2x}}{2} = k_0 \frac{(v/c)}{\sqrt{1-(v/c)^2}} \tag{6.10}$$

$$K_r = (k_{1r} + k_{2r})/2 = k_0 \sin\alpha$$

а также

$$\frac{\Omega}{k_x} = \frac{c^2}{v} = V, \qquad \frac{\omega}{K_x} = v, \qquad \frac{\Omega\omega}{k_x K_x} = Vv = c^2, \qquad \Omega\omega = c^2 k_x K_x \tag{6.11}$$

Записав, аналогично разделу 5, уравнение для идущей волны (6.8) в краткой форме

$$f = f_1 + f_2 = 2\varphi\Phi \tag{6.12}$$

получим следующие выражения для вторых частных производных $f$:

$$f_{tt} = \omega_1^2 f_1'' + \omega_2^2 f_2'' = 2(\omega^2 \varphi''\Phi + 2\omega\Omega\,\varphi'\Phi' + \Omega^2 \varphi\Phi'')$$

$$f_{xx} = k_{1x}^2 f_1'' + k_{2x}^2 f_2'' = 2(K_x^2 \varphi''\Phi + 2K_x k_x \varphi'\Phi' + k_x^2 \varphi\Phi'') \tag{6.13}$$

$$f_{rr} = k_{1r}^2 f_1'' + k_{2r}^2 f_2'' = 2(K_r^2 \varphi''\Phi)$$

где штрихами обозначены производные функций по собственным фазам. Тогда, с учётом (6.7)

$$\omega_1^2 f_1'' + \omega_2^2 f_2'' = c^2\left[(k_{1x}^2 + k_{1r}^2)f_1'' + (k_{2x}^2 + k_{2r}^2)f_2''\right] = c^2(k_1^2 f_1'' + k_2^2 f_2'') \tag{6.14}$$

то есть, функция $f$ удовлетворяет классическому волновому уравнению

$$f_{tt} = c^2(f_{xx} + f_{rr}) \tag{6.15}$$

и, исходя из $\varphi\Phi$-формы этого уравнения, получим

$$\omega^2 \varphi''\Phi + 2\omega\Omega\, \varphi'\Phi' + \Omega^2 \varphi\Phi'' = c^2\bigl(K_x^2 \varphi''\Phi + 2K_x k_x\, \varphi'\Phi' + k_x^2 \varphi\Phi'' + K_r^2 \varphi''\Phi\bigr)$$

$$(\Omega^2 - c^2 k_x^2)\varphi\Phi'' = (c^2 K_x^2 + c^2 K_r^2 - \omega^2)\varphi''\Phi$$

(6.16)

$$\Omega^2 - c^2 k_x^2 = \omega_0^2$$

где учтены: формула (6.11), равенство $\varphi''\Phi = \varphi\Phi''$ и следующее из (6.10) выражение

$$c^2 K_x^2 + c^2 K_r^2 - \omega^2 = \omega_0^2 \left[\cos^2\alpha + \left(1 - \frac{v^2}{c^2}\right)\sin^2\alpha - \frac{v^2}{c^2}\cos^2\alpha\right] \Big/ \left(1 - \frac{v^2}{c^2}\right) = \omega_0^2$$

Таким образом, аналогично одномерной волне, рассмотренной в разделе 2, для трёхмерной идущей волны на основе соотношения (6.16) и формул (6.10) имеем следующее сопоставление её параметров с характеристиками частицы:

(6.17)

$$\hbar\omega_0 \ \sim\ m_0 c^2,$$

$$\hbar\Omega = \frac{\hbar\omega_0}{\sqrt{1 - (v/c)^2}} \ \sim\ \frac{m_0 c^2}{\sqrt{1 - (v/c)^2}} = mc^2 = E,$$

$$\hbar k_x = \frac{\hbar k_0\, v/c}{\sqrt{1 - (v/c)^2}} \ \sim\ \frac{m_0 v}{\sqrt{1 - (v/c)^2}} = mv = p.$$

Отметим, что полученные соотношения справедливы для всех точек идущей волны и описывают эволюцию волнового процесса вдоль оси $Ox$ в каждой точке 3-мерного пространства.

### 7. Амплитудные закономерности

Вернёмся к уравнению (6.1) и обсудим амплитудные закономерности, которые в предыдущем разделе были оставлены без рассмотрения. В отличие от одномерной волны, все точки которой равноправны, сферическая волна имеет выделенную точку – центр сферы ($R = 0$), сопоставляемую как с началом системы координат, так и с центром моделируемой частицы. Записанная без штрихов, амплитуда волновой функции (6.1) имеет следующий вид: $|(2/R)\cos(k_0 R)|$. Она гармонически зависит от $R$ и при этом убывает пропорционально $1/R$. Таким образом, максимум амплитуды волновой функции естественным образом совпадает с центром частицы. Подчеркнём, что этот факт никак не связан с возможностью формирования волнового пакета и реализуется для произвольной *монохроматической* стоячей сферической волны (или для суперпозиции таких волн).

А.В. Горюнов

Обычно отмечают, что при $R \to 0$, амплитуда сферической волны стремится к бесконечности, и расценивают это как существенный недостаток всех таких (зависящих от $1/R$) центрально-симметричных конструкций. Для стоячей сферической волны эта проблема разрешается путём надлежащего согласования начальных фаз встречных волн. А именно, требуется, чтобы суммирующиеся сходящаяся и расходящаяся волны приходили в центр сферы в противофазе, в результате чего там бы располагался узел стоячей волны. Например, при

(7.1)
$$f_1 \leftrightarrow \cos(\omega_0 t - k_0 R) \quad \text{и} \quad f_2 \leftrightarrow \cos(\omega_0 t + k_0 R \pm \pi) = -\cos(\omega_0 t + k_0 R)$$

уравнение (6.1) примет вид

(7.2)
$$\frac{a}{R}\cos(\omega_0 t - k_0 R) - \frac{a}{R}\cos(\omega_0 t + k_0 R) = \frac{2a}{R}\sin(k_0 R)\sin(\omega_0 t) = 2ak_0 \frac{\sin(k_0 R)}{k_0 R}\sin(\omega_0 t)$$

где учтено также, что складываемые бегущие волны характеризуются собственной – не обязательно единичной – одинаковой амплитудой $a$.

При $R = 0$ дробь в правой части (7.2) равна единице и амплитуда имеет конечное значение $2ak_0$. При всех остальных значениях $R$ – дробь $|\sin(k_0 R)/k_0 R| < 1$, и стремится к нулю с ростом $R$. Таким образом, амплитуда сферической стоячей волны имеет следующую величину, конечную при всех значениях $R$:

(7.3)
$$A = 2ak_0 \left|\frac{\sin(k_0 R)}{k_0 R}\right|$$

График зависимости $A(R)$ напоминает форму волнового пакета, хотя природа такой зависимости совершенно иная.

В классической механике энергетические характеристики волны зависят от её амплитуды. В квантовой механике соответствующие характеристики частицы определяются частотой (или волновым числом) волновой функции. Формула (7.3) позволяет состыковать эти дуальные концепции, поскольку волновое число $k_0$ оказывается составной (мультипликативной) частью классической амплитуды.

Волна (7.2), кроме центрального узла имеет концентрические узловые поверхности, радиус которых определяется условием

(7.4)
$$\sin(k_0 R_n) = 0, \qquad R_n = n\pi/k_0, \qquad (n = 1, 2, \dots)$$

Отсюда следует, что в 3-мерном случае суперпозиция стоячих волн кратных частот будет иметь общие узловые поверхности. При преобразованиях Лоренца – когда происходит переход в систему отсчёта идущей волны – общие узловые поверхности останутся общими, хотя изменят сферическую форму на эллипсоидную. Устойчивой и имеющей конечную амплитуду будет и суперпозиция сферических стоячих волн с произвольным (не кратным) соотношением частот, если все они имеют общий узел в центре сферы.

**8. Заключение**

Не повторяя выводов, уже сделанных в каждом из предыдущих разделов, сформулируем несколько итоговых обобщающих замечаний.

Основным понятием данной работы является понятие *идущей* волны. Когда мы рассматриваем в некоторой системе отсчёта стоячую волну (одномерную или трёхмерную), а потом переходим в новую систему отсчёта, движущуюся относительно исходной со скоростью $v$, то тем самым и получаем в новой системе отсчёта волну, *идущую* со скоростью $v$. Отметим, что сама стоячая волна отсюда может интерпретироваться как частный случай идущей волны, имеющей скорость $v = 0$. Как было показано выше, через параметры идущей волны могут быть выражены и волновые (длины волн де Бройля и Комптона), и корпускулярные (4-вектор энергии-импульса и масса покоя) характеристики микрочастицы. В свете такого единого – волнового – описания, дуальные понятия частицы и её волновой функции также приобретают общий смысл:

1. Точечная микрочастица представляет собой некоторый волновой процесс, занимающий всё пространство. То, что обычно интерпретируют как положение покоящейся частицы, есть лишь центр сферической стоячей волны, геометрическое место, не имеющее каких-либо других принципиальных особенностей.

2. Волновая функция микрочастицы является суперпозицией волновых процессов, имеющих световую, а не сверхсветовую скорость. Тем самым, вместо абстрактных вероятностных характеристик эта функция наделяется свойствами реального физического поля, которое собственно и образует частицу, и характеризует все её свойства.

3. Наибольшую, но конечную амплитуду 3-мерная волновая функция имеет в центре сферической стоячей волны. Этот факт никак не связан с возможностью формирования волнового пакета и реализуется для произвольной (фазово-согласованной) монохроматической стоячей волны. Так же, как и для суперпозиции таких волн.

Ввиду своей элементарности и обобщённости, предложенная модель не претендует на описание какой-либо конкретной микрочастицы или какого-то определённого поля. С другой стороны, в силу тех же причин, можно полагать применимость самой идеологии данной модели для описания не только массы частицы, но и других её точечных характеристик. Так, например, фактическое отсутствие электрического заряда в предполагаемом месте его локализации снимает все проблемы, связанные с его бесконечной плотностью в точке или с его распределением внутри некой гипотетической сферы. И т. п.

Таким образом, данная работа предлагает новый взгляд на ряд уже устоявшихся понятий физики, связанных с концепцией корпускулярно-волнового дуализма, и служит обоснованию возможности принципиально иного – по сути своей, классического – волнового релятивистского подхода в этой области, основанного на концепции *идущей* волны.

Все понятия, определения и формулы, использованные в данной работе в качестве исходных, давно стали достоянием учебной литературы. Они могут быть найдены в соответствующих разделах любого Курса общей физики или в обычных справочных источниках, например [1], [2].

### Благодарности

### Литература

## Walking Wave as a Model of Particle

*A.V. Goryunov*

The concept of *walking wave* is introduced from classical relativistic positions. One- and three-dimensional walking waves considered with their wave equations and dispersion equations. It is shown that wave characteristics (de Broglie's and Compton's wavelengths) and corpuscular characteristics (energy-momentum vector and the *rest mass*) of particle may be expressed through parameters of walking wave. By that the new view on a number concepts of physic related with wave-particle duality is suggested.